\documentclass[aps, prstab, reprint, showpacs, amsmath, amssymb]{revtex4-1}

\usepackage{amsmath,amscd}
\usepackage{bm}
\usepackage[english]{babel}
\usepackage{graphicx}
\usepackage{soul}
\usepackage[usenames]{color}

\begin{document}
\title{ Radiation of a Charge in Axicon-Based Dielectric Concentrator \\ for Cherenkov Radiation }

\author{Sergey N. Galyamin}
\email{s.galyamin@spbu.ru}
\author{Andrey V. Tyukhtin}
\affiliation{Saint Petersburg State University, 7/9 Universitetskaya nab., St. Petersburg, 199034 Russia}

\date{\today}

\begin{abstract}
We propose a new type of axisymmetric dielectric target which effectively concentrates Cherenkov radiation (CR) generated in the bulk of the material into a small vicinity of focus point.
It can be called the ``axicon-based concentrator for CR''. 
A theoretical investigation of radiation field produced by a charge moving through the discussed radiator is performed for the general case where a charge trajectory is shifted with respect to the structure axis.
The idea of dielectric target with specific profile of the outer surface was presented and developed in our preceeding papers.
However, contrary to the previous configuration of such a target (which was investigated for both centered and shifted charge trajectory), the current version of the device allows efficient concentration of CR energy from relativistic particles, making this device extremely prospective for various applications.
\end{abstract}

\maketitle

\section{Introduction\label{sec:intro}}

Electromagnetic radiation emerging during an interaction between charged particle beams and various structures (homogeneous media, periodic structures etc.) were widely used for decades in various applications. 
For example, ordinary vacuum devices (clystrons, backward-wave oscillators, gyrotrons) and outstanding in power and spatiotemporal resolution of the pulse X-ray free-electron lasers are both based on beam interaction with periodic ``structure'' (be it a real structure or a specially structured external field) resulting in appropriate beam transformation and essential gain of radiation intensity.
When the medium which beam is interacted with is homogeneous Cherenkov radiation (CR) occurs even in the case of uniform motion with velocity exceeding light speed in the given medium~\cite{Ch37, TF37, Jb, Zrb, B62}.
For years, CR was intensively studied in various contexts: development of Cherenkov detectors and counters~\cite{Jb, Zrb}, CR-based microwave sources~\cite{Danos53, Coleman60}, dielectric wakefield acceleration~\cite{Ant12, JingAntipov18, OShea16}, bunch size measurement~\cite{Pot10, Kieffer18, Kieffer2020} and contemporary sources of radiation, including those for Terahertz (THz) frequencies~\cite{Takahashi00, Ant13, GTAB14, Sei17, WangAntipov17, WangAntipov2018} (note that Cherenkov-type radiation from opically rectificated laser pulses~\cite{Askaryan1962} is also considered as a convenient way to produce THz radiation in dielectric-based convertors~%
\cite{Bakunov2010, BakunovBodrov2020}). 
It is worth noting that the idea of using the CR effect for producing the radiation is not new, but in recent years it has been essentially imroved.
If a bunch used for generation is already of proper quality (this is typically so for bunches produced by modern accelerators) then an extraordinary peak power can be potentially obtained in relatively simple dielectric-loaded structures~\cite{OShea16}.
Non-invasive bunch diagnostics based on prolonged dielectric targets of complicated shape is another modern area for CR applications~\cite{Pot10, Kieffer18, Kieffer2020} which possess several advantages compared to traditional schemes based on transition or diffraction radiation.

The last mentioned area involves the need to calculate CR produced by dielectric target with seveal boundaries and edges which is marginally possible to do rigoriusly.
To reslove this issue with reliable accuraccy we have been developing for several recent years an original combined approach based on certain ``etalon'' problem, ray-optics laws and Stratton-Chu formulas~\cite{BTG13, BGT15, GTV17, GTV18, TGV19, TVGB19, GVT19, GVT19E, TBGV20arxiv, TGVGrig20arxiv}.
It is worth noting that this approach has been approved by direct comparison between its results and resuts of numerical simulations in COMSOL Multiphysics~\cite{GTV18, TBGV20arxiv}.

Several paper from this list closely relate to the present paper because they dealt with the axisymmetric dielectric target -- ``dielectric concentrator for CR'' -- focusing the majority of generated CR in a small vicinity of a predetermined point (focus) without any additional lenses or mirrors~\cite{GT14, GTV17, GTV18, GVT19, GVT19E}.
While possibilities of this concentrator for radiation intensity enhancement, beam position and velocity measurements are rather attractive, an essential disadvantage is that this target has moreless convenient dimensions for relatively slow charged particles only.
Therefore it would be of considerable practical importance to eliminate the mentioned shortcoming and allow relativistic charged particle to concentrate produced CR. Solution of this problem is the main goal of the present paper.

We propose here an ``axicon-based concentrator for CR'' - a new type of axisymmetric dielectric target which concentrates the main portion of CR into a focus point.
Contrary to the previous ``single-refraction'' configuration~\cite{GT14}, this target uses one reflection and one refraction of CR rays.
For reflection it is convenient to use a hollow conical target (axicon) in the geometry investigated separately in recent paper~\cite{TGVGrig20arxiv}.
It is important here that the cone apex angle can be adjusted so that CR rays will form paraxial beam with respect to the charge trajectory after the reflection for arbitrary angle of incidence (i.e., for arbitrary charge velocity).
Therefore the discussed concentrator can be designed for effective focusing of CR from chaged particle bunches with arbitrary velocity including relativistic velocities which are of most practical interest.
Moreover, we investigate here the effect of charge shift from the symmetry axis, similarly to the analogous investigation for ``single-refraction'' concentrator~\cite{GVT19, GVT19E}. 


%
\begin{figure*}[t]
\centering
\includegraphics[width=0.85\linewidth]{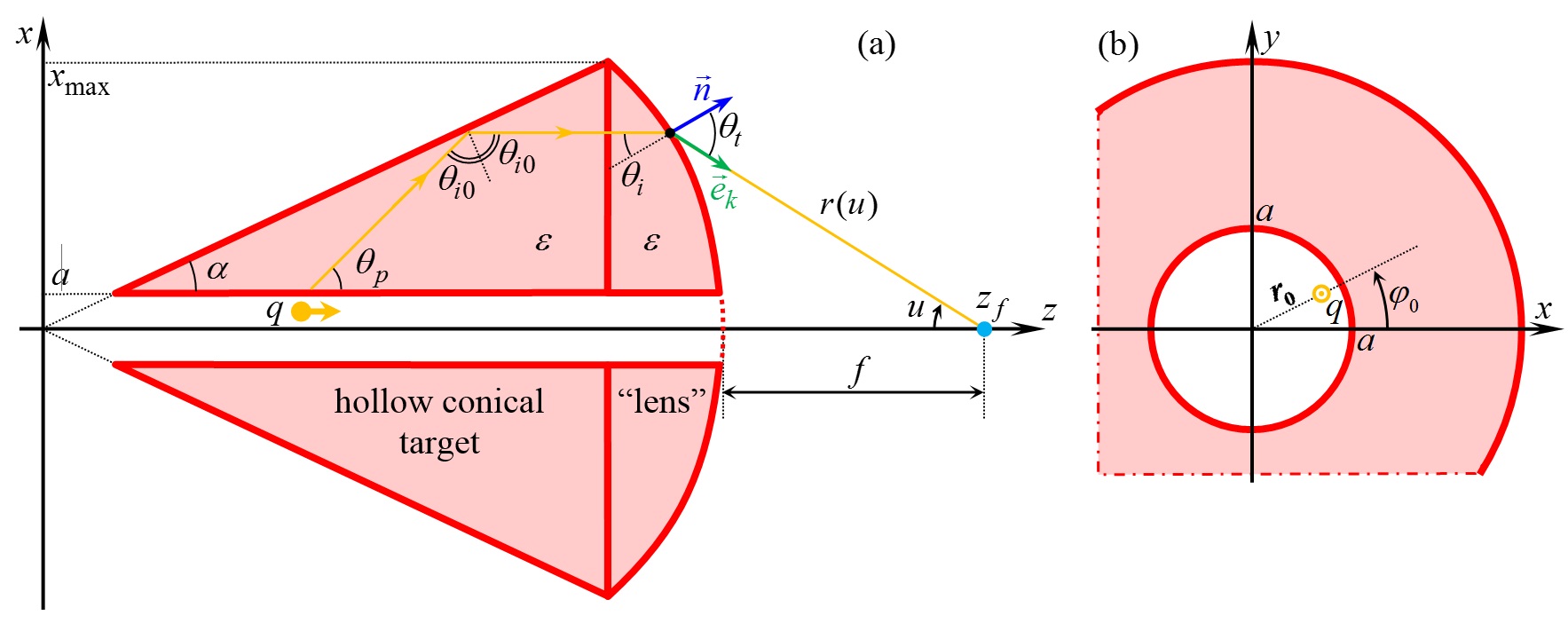}
\caption{\label{fig:geom}%
Geometry of the problem and main notations.
(a)
$ (zx) $%
-cut of the ``axicon-based concentrator for CR'': a hollow ``lens'' is attached to the output surface of a hollow conical target having its apex facing the incident charged particle bunch. Both revolution bodies are made from the same dielectric material with permittivity $\varepsilon$. The outer profile of the ``lens'' is hyperbolic and is determined by the function
$ r( u ) $.
A point charge
$ q $
moves along the straight trajectory shifted with respect to the
$ z $%
-axis. 
Depicted parameters are discussed in the text.
(b) $ (xy) $%
-cut of the target (the channel radius is enlarged for convenience) and the position of the charge shifted trajectory.
}
\end{figure*}

\section{Problem formulation\label{sec:problemstat}}

Figure~%
\ref{fig:geom}
shows the geometry of the problem under investigation.
Note that along with Cartesian frame
$ ( x, y, z ) $,
corresponding cylindrical frame
$ ( \rho, \varphi, z ) $
is introduced.
A point charge
$ q $
moves with a constant velocity
$ \upsilon = \beta c $
along straight trajectory inside the channel in the axisymmetric dielectric target with permittivity
$ \varepsilon $
and permeability
$ \mu = 1 $ (for convenience).
It is supposed here that condition for CR generation is fulfilled, i.e.
$ \sqrt{\varepsilon} \beta > 1 $.
Position of charge trajectory in $xy$-plane is determined by
$ r_0 $
and
$ \varphi_0 $%
, see Fig.~%
\ref{fig:geom}~%
(b).

The target consists of two ``glued'' bodies of revolution: a hollow cone and a hollow ``lens''. The cone is determined by its apex angle $\alpha$ while cylindrical coordinates
$ \rho = \rho_0 $,
$ z = z_0 $
of the outer profile of the ``lens'' are determined as follows (note that this profile can be deduced using the same considerations as in~\cite{GT14, GVT19}):
\begin{equation}
\label{eq:concsurf}
\begin{aligned}
\rho_0 ( u ) &= r( u ) \sin ( u ), \\
z_0 ( u ) &= z_f - r( u ) \cos ( u ),
\end{aligned}
\end{equation}
where
\begin{equation}
\label{rU}
r( u )
=
f ( \sqrt{\varepsilon} - 1 ) \left[ \sqrt{\varepsilon} \cos ( u ) - 1 \right]^{ -1 },
\end{equation}
$ f $
is a ``focal'' parameter
The maximum transverse size of the target
$ x_{ \max } $
determines the maximum angle
$ u_{ \max } $,
the minimum angle
$ u_{ \min } $
is determined by the channel radius
$ a $.
Total length of the target including the imaginary ``nose'' of the cone is $ z_f - f $, where $ z_f = x_{\max} (\cot \alpha + \cot u_{\max} ) $.
It should be underlined that the surface~\eqref{rU} is typically designed so that a beam of rays parallel to $z$-axis converges to exact focus $z=z_f$ after the refraction.
Schematic propagation of one such ray through the target to the focus is shown in Fig.~\ref{fig:geom}.
It is of essential importance that for ``axicon-based concentrator for CR'' we have \emph{two} independent parameters for design, charge velocity $\beta$ and cone angle $\alpha$ .
This means, for example, that we can adjust the cone angle $\alpha$ so that CR rays will be parallel to $z$-axis for \emph{arbitrary} charge velocity including relativistic velocities which are of most practical importance. This point will be discussed again below.

Our further analysis is at first based on analytical solution of the corresponding ``etalon'' problem (see~\cite{GVT19, GVT19E}) allowing determination of initial CR rays inside the bulk of dielectric.
Important note to this solution (which is an ifinite sum of ``harmonics'') is the following: the phase of all summands can be written as
\begin{equation}
\label{phase}
\exp \left[ i k_0 \beta^{-1}
\left(
z + \sqrt{ \varepsilon \beta^2 - 1 } \rho
\right)  \right],
\end{equation}
i.e. in the same manner as in the symmetric case
$ r_0 = 0 $ (here $ k_0 = \omega / c $, $ \omega $ is a frequency, $ c $ is a light speed in vacuum).
Therefore all results concerning the structure of CR rays inside the hollow conical target for a symmetric case (see~\cite{TGVGrig20arxiv}) are fully applicable to the present problem. In particular, we should consider only the ``main'' wave of two analyzed in~\cite{TGVGrig20arxiv} because the ``lens'' surface~\eqref{rU} has been designed for concentration of paraxial CR rays which correspond to this wave (it is worth noting that in the case of flat output surface it is this wave which is responsible for ``Cherenkov spotlight'' effect).
In our case, propagation of this wave is the same untill ray refraction at the output surface because now the ``lens'' is attached there. 
This last refraction  can be analyzed similarly to the case of a ``single-refraction'' concentrator. 


\section{Stratton-Chu formalism\label{sec:SChu}}

According to our method we utilize the Stratton-Chu formulas~%
\cite{SChu39, Fradb}
to calculate CR exiting the target.
Recall that these integral formulas give exact result if tangential electric and magnetic fields are determined exactly at the surface of the integration (the aperture
$ S _a $%
).
In this paper we use the form of these formulas from~%
\cite{Fradb}
(see also our papers~%
\cite{GT17, TGV19, TVGB19, GVT19}%
) with the outer surface of the target~%
\eqref{eq:concsurf}
as the aperture:
\begin{equation}
\label{SChu}
\begin{aligned}
&4 \pi
\vec{ E }_{ \omega }
=
\int\nolimits_{ S_a }
\left\{
\vphantom{ \frac{ i }{ k_0 } }
i k_0
\left[
\vec{ n }, \, \vec{ H }_{ \omega }^a
\right]
\psi
+
\right. \\
&+
\left.
\frac{ i }{ k_0 }
\left(
\left[
\vec{ n }, \, \vec{ H }_{ \omega }^a
\right], \, \vec{ \nabla }
\right)
\!
\vec{ \nabla }
\psi
+
\left[
\left[
\vec{ E }_{ \omega }^a, \, \vec{ n }
\right], \, \vec{ \nabla } \psi
\right]
\right\}
d \Sigma,
\end{aligned}
\end{equation}
where
$ \psi $
is a Green function,
\begin{equation}
\label{eq:Green}
\begin{aligned}
\psi
&= \left. \exp \left( i k_0 \tilde{ R } \right) \right/ \tilde{ R }, \\
\tilde{ R }
&=
\sqrt{ ( x - x_0 )^2 + ( y - y_0 )^2 + ( z - z_0 )^2 },
\end{aligned}
\end{equation}
$ d \Sigma $
is a surface element of
$ S_a $.
As follows from Eq.~%
\eqref{SChu},
electromagnetic (EM) field outside the target is determined by the tangential electric and magnetic fields at the aperture
$ S_a $.
We utilize the following parametrization of the Cartesian coordinates of the aperture via angles
$ u $
and
$ \varphi $:
\begin{equation}
\label{x0y0param}
x_0( u, \varphi ) = \rho_0( u ) \cos \varphi,
\quad
y_0( u, \varphi ) = \rho_0( u ) \sin \varphi,
\end{equation}
while
$ z_0 ( u, \varphi ) $
is given by~%
\eqref{rU}
together with
$ \rho_0 ( u ) $.
In order to calculate the parameters of the surface it is convenient to use the tensor formalism by V.A.~Fock~%
\cite{Fockb, GTV17}
and determine the metric tensor of the surface
$ g $.
Thus, for the elementary square of the surface we obtain
$ d \Sigma = \sqrt{ g( u ) } d u d \varphi $,
where
\begin{equation}
\label{eq:sqrtg}
\sqrt{ g(u) }
=
\frac{
r^2( u ) \sin u \sqrt{ 1 - 2 \sqrt{ \varepsilon } \cos u + \varepsilon } }{\sqrt{\varepsilon} \cos u - 1}.
\end{equation}
The components of the external unit normal
$ \vec{ n } $
are:
\begin{equation}
\label{eq:normal}
\begin{aligned}
n_{ \rho }
&=
\frac{ \sin u }
{ \sqrt{ 1 + \sqrt{ \varepsilon } \cos u + \varepsilon } }, \\
n_{ z }
&=
\frac{ \sqrt{\varepsilon} - \cos u }
{ \sqrt{ 1 + \sqrt{ \varepsilon } \cos u + \varepsilon } }.
\end{aligned}
\end{equation}

In order to find the fields
$ \vec{ E }_{ \omega }^a $
and
$ \vec{ H }_{ \omega }^a $,
we use the same approved method as in papers~%
\cite{GTV18, GVT19}.
Moreover, corresponding ``etalon'' problem is the same as in \cite{GVT19}.
The solution of the ``etalon'' problem should be compemented by the analysis of the ``main'' CR wave reflection at the cone generatrix, this has been done in~\cite{TGVGrig20arxiv}.
In particular, the way of this ray is the following.
It incidents the boundary formed by cone generatrix at the angle
\[
\theta_{ i 0 } = \pi / 2 + \alpha - \theta_p,
\quad
\text{where}
\quad
\theta_p = \arccos \left[ \frac{ 1 }{ \sqrt{\varepsilon} \beta } \right], 
\]
see Fig.~%
\ref{fig:geom}%
, reflected at the same angle and some portion of it is refracted at the angle
\[
\theta_{ t 0 } = \arcsin ( \sqrt{\varepsilon} \sin \theta_{ i 0 } )
\]
with respect to the generatrix normal.
Then the reflected wave propagates to the cone base and passes through the cone base flat surface at the angle 
\[
\theta_i^{ \prime } = 2 \alpha - \theta_p
\]
with respect to $z$-axis. 
When $ \theta_i^{ \prime } = 0 $ the rays form a parallel beam before the ``lens'', this is the case where all the rays converge exactly to the focus point. 
This is very important feature of the discussed ``axicon-based concentrator'' that we have two parameters to obtain parallel beam of rays: $\alpha$ and $\beta$.
Most convenient way is to adjust $\alpha$ in accordance to the given charged particle velocity $\beta$, i.e. to chose
\begin{equation}
\alpha = \frac{1}{2} \theta_p =\frac{1}{2}\arccos\left(\frac{1}{\sqrt{\varepsilon}\beta}\right).
\end{equation}
Also this posiibility is the most distinguishing point compared to the ``single-refraction'' concentrator \cite{GT14,GVT19,GVT19E} which can be designed for single charge velocity only.

Due to the asymmetry caused by charge shift, CR of both polarizations is generated. 
``Parallel'' polarization (%
$ \parallel $%
) contains components
$ E_{ z \omega } $,
$ E_{ r \omega } $
and
$ H_{ \varphi \omega } $.
Corresponding Fresnel reflection coefficient (responsible for the reflection at the cone generatrix) is
\begin{equation}
R_{ \parallel }
=
\frac{ \cos \theta_{ i 0 } - \sqrt{ \varepsilon } \cos \theta_{ t 0 } }
{ \cos \theta_{ i 0 } + \sqrt{ \varepsilon } \cos \theta_{ t 0 } },
\label{Rpar}
\end{equation}
while Fresnel transmision coefficient (for the transmission at the ``lens'' output surface) is
\begin{equation}
\label{Tpar}
T_{ \parallel }
=
\frac{ 2 \cos \theta_{ i } }{ \cos \theta_{ i } + \sqrt{ \varepsilon } \cos \theta_{ t } }.
\end{equation}
``Orthogonal'' polarization (%
$ \bot $%
) contains components
$ H_{ z \omega } $,
$ H_{ \rho \omega } $
and
$ E_{ \varphi \omega } $.
Corresponding Fresnel coefficients are:
\begin{align}
R_{ \bot }
&=
\frac{ \sqrt{ \varepsilon } \cos \theta_{ i 0 } - \cos \theta_{ t 0 } }
{ \sqrt{ \varepsilon } \cos \theta_{ i 0 } + \cos \theta_{ t 0 } },
\label{Rort} \\
T_{ \bot }
&=
\frac{ 2 \sqrt{ \varepsilon } \cos \theta_{ i  } }{ \sqrt{ \varepsilon } \cos \theta_{ i  } + \cos \theta_{ t  } }.
\label{Tort}
\end{align}
The angle of incidence
$ \theta_{ i } $
can be obtained out of Snell's law,
$ \sqrt{ \varepsilon } \sin \theta_{ i  } = \sin \theta_{ t } $,
while $ \theta_{ t } $ is calculated a bit later.

Based on above considerations, we can write out transmitted field at the outer surface of the aperture which is determined via the component of the field being orthogonal to the plane of incidence, i.e.
$ H_{ \varphi \omega } $
for
$ \parallel $%
-polarization and
$ E_{ \varphi \omega } $
for
$ \bot $%
-polarization.
First, at the inner side of the aperture we have (see~\cite{GVT19, GVT19E,TGVGrig20arxiv} for details):
\begin{equation}
\label{Hphiappr}
\begin{aligned}
&H_{ \varphi \omega }^{ a- }
=
\frac{ q \omega e^{ i k_0 \sqrt{ \varepsilon } \left( z_0 \cos \theta_i^{ \prime } + \rho_0 \sin \theta_i^{ \prime } \right) - \frac{ 3 \pi i }{ 4 }  } }
{ i \pi \upsilon^2 \gamma^2 }
\sqrt{ \frac{ 2 }{ \pi \rho_0 s } }
{\times} \\
&{\times}
\frac{ i k_0 }{ s^2 }
\!\!
\left\{
- \varepsilon s
I_0 ( r_0 \sigma_0 )
\tilde{ A }_0^{ ( E2 ) }
{+}
2
\sum\limits_{ \nu = 1 }^{ \infty }
I_{ \nu } ( r_0 \sigma_0 )
e^{ \frac{ i \pi ( 1 {-} \nu ) }{ 2 } }
{\times}
\right. \\
&\left.
\vphantom{ \sum\limits_{ \nu = 1 }^{ \infty } }
\times
\cos ( \nu \varphi )
\left[
\varepsilon
\tilde{ A }_{ \nu }^{ ( E2 ) }
\left( i s - \frac{ 1 }{ 2 \rho_0 } \right)
-
\frac{ i \nu }{ \beta \rho_0 }
\tilde{ A }_{ \nu }^{ ( H2 ) }
\right]
\right\},
\end{aligned}
\end{equation}
\begin{equation}
\label{Ephiappr}
\begin{aligned}
&E_{ \varphi \omega }^{ a- }
=
\frac{ q \omega e^{ i k_0 \sqrt{ \varepsilon } \left( z_0 \cos \theta_i^{ \prime } + \rho_0 \sin \theta_i^{ \prime } \right) - \frac{ 3 \pi i }{ 4 }  } }
{ i \pi \upsilon^2 \gamma^2 }
\sqrt{ \frac{ 2 }{ \pi \rho_0 s } }
{\times} \\
&{\times}
\frac{ i k_0 }{ s^2 }
2
\sum\limits_{ \nu = 1 }^{ \infty }
I_{ \nu } ( r_0 \sigma_0 )
e^{ \frac{ i \pi ( 1 {-} \nu ) }{ 2 } }
{\times} \\
&{\times}
\sin ( \nu \varphi )
\left[
i
\tilde{ A }_{ \nu }^{ ( H2 ) }
\left( i s - \frac{ 1 }{ 2 \rho_0 } \right)
-
\frac{ \nu }{ \beta \rho_0 }
\tilde{ A }_{ \nu }^{ ( E2 ) }
\right],
\end{aligned}
\end{equation}
%
%
\begin{equation}
\label{eq:AnuE2H2}
\tilde{ A }_{ \nu }^{ ( E2 ) }
{=}
\tilde{ A }_{ \nu }^{ ( E1 ) }
\frac{ I_{ \nu } }{ H_{ \nu } }
+
\frac{ K_{ \nu } }{ H_{ \nu } },
\quad
\tilde{ A }_{ \nu }^{ ( H2 ) }
{=}
\tilde{ A }_{ \nu }^{ ( H1 ) }
\frac{ I_{ \nu } }{ H_{ \nu } },
\end{equation}
\begin{equation}
\label{eq:AnuE1}
\begin{aligned}
&\tilde{ A }_{ \nu }^{ ( E1 ) }
{=}
\frac{ 1 }{ \Delta_{ \nu } H_{ \nu }^2 }
\left\{
-\left[
\nu ( \beta a )^{ -1 } I_{ \nu } ( \sigma_0^2 + s^2 )
\right]^2
H_{ \nu }^2 K_{ \nu } I_{ \nu }^{ -1 } \right. + \\
&{+} 
\left. \left[
\sigma_0^2 s \mu H_{ \nu }^{ \prime } I_{ \nu } {+}
s^2 \sigma_0 I_{ \nu }^{ \prime } H_{ \nu }
\right] 
\left[
\sigma_0^2 s \varepsilon H_{ \nu }^{ \prime } K_{ \nu } {+}
s^2 \sigma_0 K_{ \nu }^{ \prime } H_{ \nu }
\right]
\right\},
\end{aligned}
\end{equation}
\begin{equation}
\label{eq:AnuH1}
\begin{aligned}
&\tilde{ A }_{ \nu }^{ ( H1 ) }
{=}
\frac{ \nu I_{ \nu } ( \sigma_0^2 + s^2 ) }{ i \beta a \Delta_{ \nu } H_{ \nu } }
\left\{
\left[
\sigma_0^2 s \varepsilon H_{ \nu }^{ \prime } K_{ \nu } {+}
s^2 \sigma_0 K_{ \nu }^{ \prime } H_{ \nu }
\right] - \right. \\
&\left. -
K_{ \nu } I_{ \nu }^{ -1 }
\left[
\sigma_0^2 s \varepsilon H_{ \nu }^{ \prime } I_{ \nu } {+}
s^2 \sigma_0 I_{ \nu }^{ \prime } H_{ \nu }
\right]
\right\},
\end{aligned}
\end{equation}
\begin{equation}
\label{eq:Det}
\begin{aligned}
&\Delta_{ \nu }
=
\left[
\nu ( \beta a )^{ -1 }
I_{ \nu }
\left( \sigma_0^2 + s^2 \right)
\right]^2
- \\
&-
\frac{
\left[
\sigma_0^2 s \varepsilon H_{ \nu }^{ \prime } I_{ \nu } {+}
s^2 \sigma_0 I_{ \nu }^{ \prime } H_{ \nu }
\right]
\!
\left[
\sigma_0^2 s \mu H_{ \nu }^{ \prime } I_{ \nu } {+}
s^2 \sigma_0 I_{ \nu }^{ \prime } H_{ \nu }
\right]
}
{ H_{ \nu }^2 },
\end{aligned}
\end{equation}
\begin{equation}
\label{eq:IHprimeIH}
\begin{aligned}
I_{ \nu } &\equiv I_{ \nu }( a \sigma_0 ),
\quad
H_{ \nu } \equiv H_{ \nu }^{ ( 1 ) }( a s ),
\\
I_{ \nu }^{ \prime } &\equiv \left. \frac{ d I_{ \nu }( \xi ) } { d \xi }
\right|_{ \xi = a \sigma_0 },
\quad
H_{ \nu }^{ \prime } \equiv \left. \frac{ d H_{ \nu }^{ ( 1 ) }( \xi ) }{ d \xi }
\right|_{ \xi = a s },
\end{aligned}
\end{equation}
\begin{equation}
\label{eq:KprimeK}
K_{ \nu } \equiv K_{ \nu }( a \sigma_0 ),
\quad
K_{ \nu }^{ \prime } \equiv \left. \frac{ d K_{ \nu }( \xi ) } { d \xi }
\right|_{ \xi = a \sigma_0 }.
\end{equation}
The transmitted fields at the outer surface of the aperture are
\begin{equation}
\label{eq:EphiHphi}
\begin{aligned}
H_{ \varphi \omega }^{ a }
&=
T_{ \parallel }
H_{ \varphi \omega }^{ a- },
\quad
\vec{ E }_{ \omega }^{ \parallel }
= H_{ \varphi \omega }^{ a }
\left[
\vec{ e }_{ \varphi }, \, \vec{ e }_k
\right], \\
E_{ \varphi \omega }^{ a }
&=
T_{ \bot }
E_{ \varphi \omega }^{ a- },
\quad
\vec{ H }_{ \omega }^{ \bot }
= E_{ \varphi \omega }^{ a }
\left[
\vec{ e }_k, \, \vec{ e }_{ \varphi }
\right],
\end{aligned}
\end{equation}
therefore
\begin{equation}
\label{eq:ErhozHrhoz}
\begin{aligned}
E_{ \rho \omega }^{ a }
&=
H_{ \varphi \omega }^{ a }
e_{ k z },
\quad
E_{ z \omega }^{ a }
=
- H_{ \varphi \omega }^{ a }
e_{ k \rho },
\\
H_{ \rho \omega }^{ a }
&=
- E_{ \varphi \omega }^{ a }
e_{ k z },
\quad
H_{ z \omega }^{ a }
=
E_{ \varphi \omega }^{ a }
e_{ k \rho }.
\end{aligned}
\end{equation}
Here the unit vector of the transmitted wave
$ \vec{ e }_k $
is the following:
\begin{equation}
\label{eq:ek}
\begin{aligned}
e_{ k \rho } &= n_{ \rho } \cos \theta_t - n_z \sin \theta_t, \\
e_{ k z } &= n_{ \rho } \sin \theta_t + n_z \cos \theta_t.
\end{aligned}
\end{equation}
The last point is the refraction angle $ \theta_t $ which can be calculated using the phase term from~\eqref{Hphiappr}, \eqref{Ephiappr} and metric tensor of the outer surface \eqref{rU} (see~\cite{Fockb, GTV17} for details):
\begin{equation}
\label{eq:sintheta}
\begin{aligned}
&\sin \theta_t
=
\frac{ \sqrt{ \varepsilon } } { \sqrt{ 1 - 2 \sqrt{ \varepsilon } \cos u + \varepsilon } } \times \\
&\times
\left[ \sin \left( 2 \alpha - \theta_p + u \right) - 
\sqrt{ \varepsilon } \sin \left( 2 \alpha - \theta_p \right) \right].
\end{aligned}
\end{equation}
Now all the things needed for the evaluation of the integral~%
\eqref{SChu}
are ready.

%
\begin{figure*}
\centering
\includegraphics[width=0.99\linewidth]{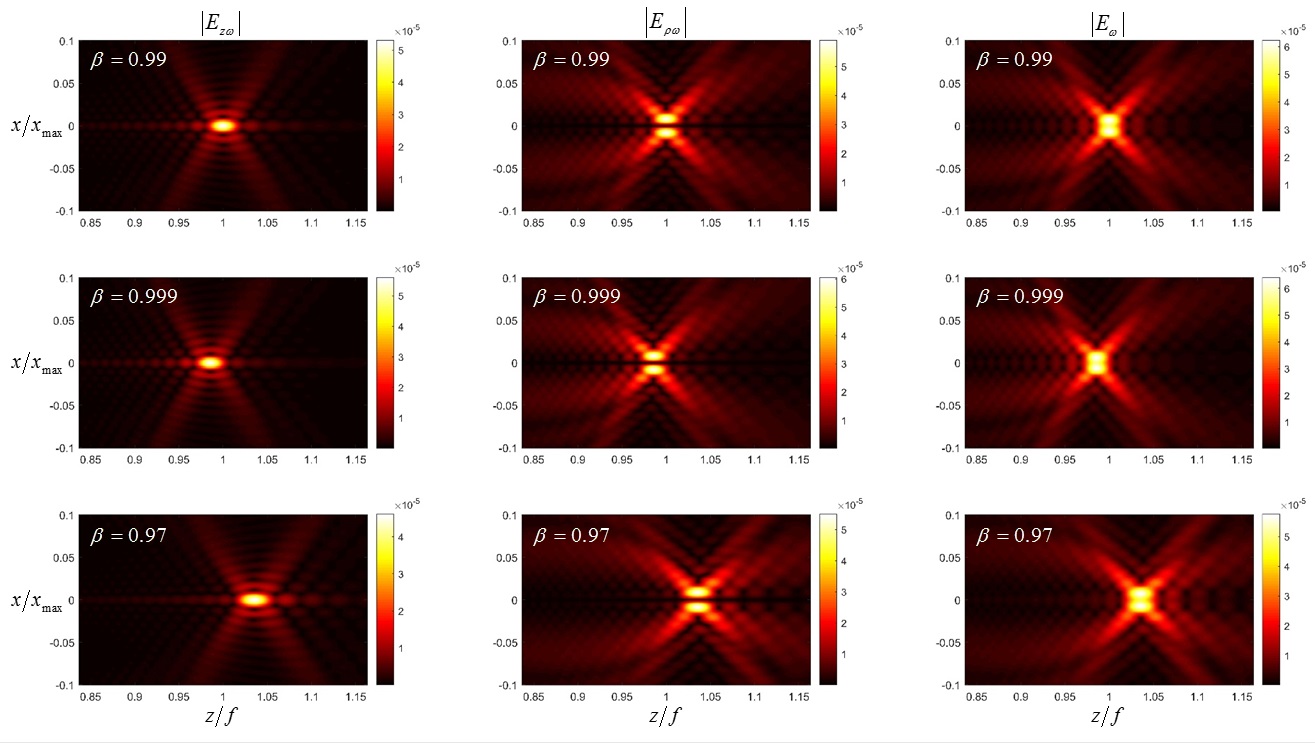}
\caption{\label{EZXsymm}%
Field distribution over $zx$-plane for symmetrical case $ r_0 = 0 $ and various charge velocities $ \beta $: top row for $\beta = 0.99$, middle row for $\beta = 0.999$, bottom row for $\beta = 0.97$. Concentrator has been designed with $\varepsilon = 4$ for $\beta = 0.99$ with the following parameters: $x_{\max} = f = 23.9$cm, $ a = c / \omega = 0.047$cm, $\alpha = 30^{\circ}$, $\omega = 2\pi\cdot 100$GHz. Field units are $\mathrm{Vms}^{-1}$.
}
\end{figure*}
%

\section{\label{sec:num}Numerical results}

Here we present results of EM field calculation in the area outside the target (mainly, near the focal point which is of the most interest) using Eq.~%
\eqref{SChu}.
The limits of integration over
$ \varphi $
are $ \left( 0, 2 \pi \right) $,
while the limits of integration over
$ u $ ($ u_{ \min } $ and $ u_{ \max } $ ) are determined by transverse dimensions of the concentrator.
A numerical code was realized in MATLAB with the use of Parallel Computing Toolbox for evaluation of integrals~%
\eqref{SChu}.

%
\begin{figure*}
\centering
\includegraphics[width=0.9\linewidth]{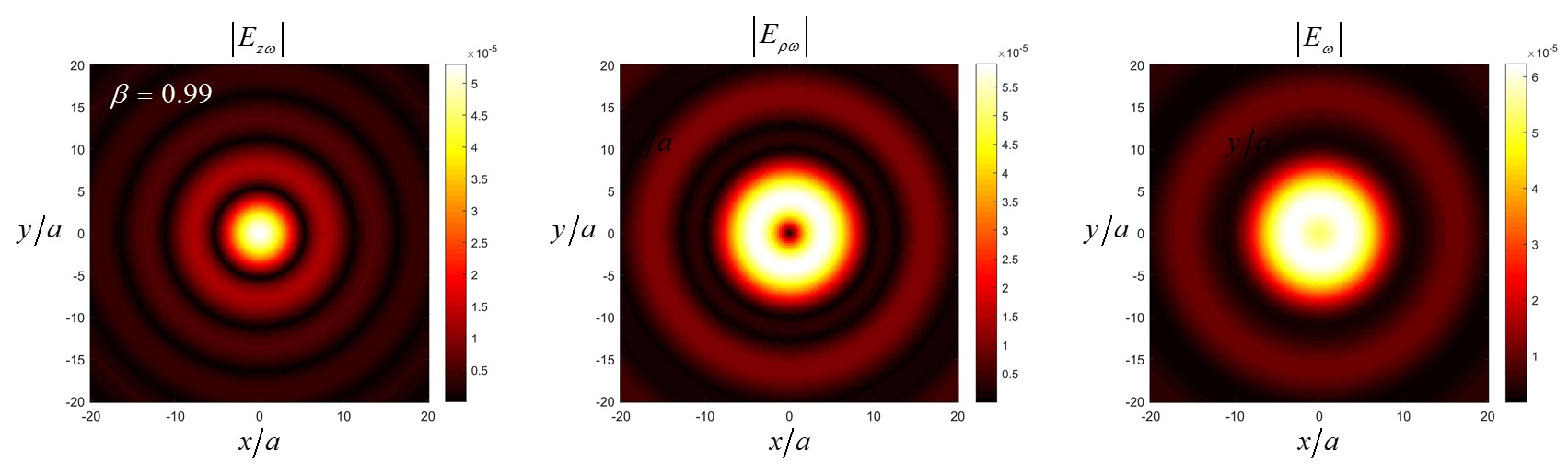}
\caption{\label{EXYsymm}%
Two-dimensional field distribution over $xy$-plane for symmetrical case $ r_0 = 0 $, $ \beta = 0.99 $ and $z = z_f$ (focal plane). 
Other parameters are the same as in Fig~\ref{EZXsymm}.
}
\end{figure*}
%

Figure~%
\ref{EZXsymm}
illustrates distribution of absolute values of
$ E_{ z \omega } $,
$ E_{ \rho \omega } $
and total field
$ E_{ \omega } $
over $zx$-plane for $ r_0 = 0 $.
One can see that transverse field
$ E_{ \rho \omega } $
is negligible near
$ z $%
-axis, which is natural due to the symmetry considerations, while longitudinal field
$ E_{ z \omega } $
is dominant there.
It is worth noting that maximum values of transverse and longitudinal fields are practically equal.
This is more favorable compared to the ``single-refraction'' concentrator \cite{GVT19}.
As one can see, position of the focal spot changes with the change in $\beta$ but for $\beta=0.99$ (the velocity which concentrator is designed for) the spot is exactly for $z=z_f$ which is natural.
Figure~%
\ref{EXYsymm}
illustrates similar distribution over $xy$ plane.
One can conclude that this distribution is very close to the case of ``single-refraction'' concentrator.

\begin{figure*}
\centering
\includegraphics[width=0.99\linewidth]{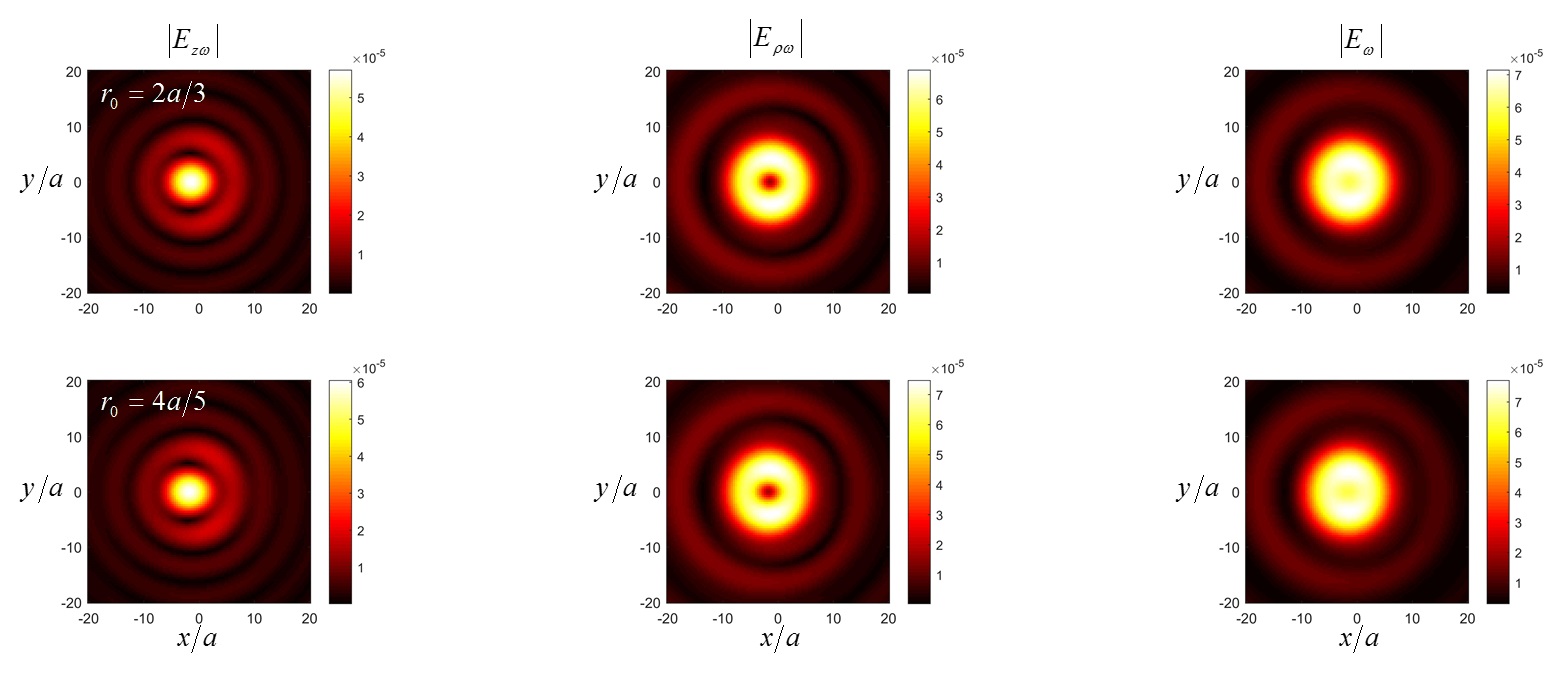}
\caption{\label{EXYasymm}%
Two-dimensional field distribution over $xy$-plane for two asymmetric cases $ r_0 = 3 a / 4 $ and $ r_0 = 4 a/5 $, $ \beta = 0.9 $ and $z = z_f$ (focal plane). 
Other parameters are the same as in Fig~\ref{EZXsymm}.}
\end{figure*}

The field distribution for the case with a charge shifted from the
$ z $%
-axis ($\varphi_0 = 0$, $r_0 \ne 0$) is shown in Fig.~%
\ref{EXYasymm}
for the focal plane
$ z = z_f $.
For calculations, two asymmetric modes (which are generated in the asymmetic case) were taken into account.

It should be underlined that in this paper we are mostly considering concentrators for relativistic charges, $ \beta \to 1 $, which are most interesting for particle accelerator applications and which can be effectively used with this device. 
As it was discussed in~\cite{GVT19}, EM field of asymmetric modes is proportional to the term
$ I_{ \nu } ( r_0 \sigma_0 ) $, $\nu>1$,
while 
$ \sigma_0 \sim \sqrt{ 1 - \beta^2 } $.
Therefore asymmetry in the EM field produced by both concentrators (``single-refraction'' and ``axicon-based'') will be essentially weaker for relativistic particles compared to nonrelativistic ones.
However, since ``single-refraction'' concentrator is only convenient for relatively slow particles, asymmetry issues are essential for its operation.
On the contrary, ``axicon-based'' concentrator is most suitable for relativistic charges and the shift of the trajectory will affect it weaker. 
To illustrate this, we chose a moderate relativistic velocity $\beta = 0.9$ and consider large offsets.
As one can see from Fig.~%
\ref{EXYasymm}, even if the charge trajectory is close to the channel wall only small asymetry in the longitudinal field can be observed: the fied maximum is slightly shifted to positive $x$.
A small redistribution of the transverse field in $y$-direction can be also observed.
Again, magnitudes of transverse and longitudinal fields are of the same order.
%
%

\section{\label{sec:concl}Conclusion}

In the present paper, we have presented analytical and numerical investigation of the EM radiation produced by a point charge moving through the new type of dielectric concentrator for Cherenkov radiation -- ``axicon-based concentrator for CR''.
The first advantage of this target compared to the ``ordinary'' one (``single-refraction'' concentrator, see~\cite{GT14,GVT19,GVT19E}) is the possibility to chose an appropriate cone angle to realize concentration of CR produced by charged particle bunches with arbitrary velocity.
This fact can be of essential importance for various application of the considered target with relativistic charged particles from modern accelerators which are mainly relativistic.
For example, strong longitudinal electric field inside the focal spot (transverse field is zero here) can be possibly used for bunch modulation.
It is worth noting that that longitudinal electric field is of the same order as the transverse field which was not the case for the ``ordinary'' device.
Another feature is that generation of asymmetric modes (which disturb the symmetric field distribution in the focal plane) can be much less important for this target even for extremely large offsets of the bunch trajectory.

\section{Acknowledgments}

This work was supported by Russian Science Foundation, grant No.~18-72-10137.


%

\end{document}